\documentclass[showpacs,twocolumn,aps,prb]{revtex4}
\usepackage{mathrsfs}
\usepackage{amsmath}
\usepackage{amsfonts}
\usepackage{amssymb}
\usepackage{amsthm}
\usepackage{graphicx}
\usepackage{color}
\usepackage{epsfig}
\usepackage{cancel}
\usepackage{bm}
\usepackage{sidecap}

\providecommand{\U}[1]{\protect\rule{.1in}{.1in}}
\usepackage{ulem}
\begin{document}
\title{Topological Anderson Insulators in Systems without Time-Reversal
Symmetry}
\author{Ying Su$^{1,2}$}
\author{Y. Avishai$^{3,4}$}
\author{X. R. Wang$^{1,2}$}
\email[corresponding author: ]{phxwan@ust.hk}
\affiliation{$^{1}$Physics Department, The Hong Kong University of
Science and Technology, Clear Water Bay, Kowloon, Hong Kong}
\affiliation{$^{2}$HKUST Shenzhen Research Institute, Shenzhen 518057,
China}
\affiliation{$^{3}$Department of Physics, Ben-Gurion University of
the Negev, Beer-Sheva, Israel}
\affiliation{$^{4}$Department of Physics, NYU-Shanghai University, Shanghai, China}
\date{\today}
\begin{abstract}
Occurrence of topological Anderson insulator (TAI) in HgTe quantum well
suggests that when time-reversal symmetry (TRS) is maintained, the pertinent
topological phase transition, marked by re-entrant $2e^2/h$ quantized
conductance contributed by helical edge states, is driven by disorder.
Here we show that when TRS is broken, the physics of TAI becomes even richer.
The pattern of longitudinal conductance and nonequilibrium local current
distribution displays novel TAI phases characterized by nonzero Chern numbers,
indicating the occurrence of multiple chiral edge modes.
Tuning either disorder or Fermi energy (in both topologically trivial and
nontrivial phases), drives transitions between these distinct TAI phases,
characterized by jumps of the quantized conductance from
$0$ to $e^2/h$ and from $e^2/h$ to $2e^2/h$. 
An effective medium  theory based on the Born approximation
yields an accurate description of different TAI phases in parameter space.
\end{abstract}

\pacs{73.43.Nq, 72.15.Rn, 72.25.-b, 85.75.-d}
\vspace{-0.2in}
\maketitle

\section{Introduction}

The quest for understanding novel electronic
properties of topological insulators (TIs) stirred intense experimental
and theoretical studies \cite{T11,T1,T2,T3,T4,T5,T6,T7,T9}.
Observations of quantized conductance at $2e^2/h$ in time reversal
invariant TIs \cite{T3,T4} and at $e^2/h$ in magnetically
doped TIs \cite{T9} confirm the occurrence of the quantum
spin Hall (QSH) and the quantum anomalous Hall (QAH) effects.
A well accepted paradigm is that the main distinction of TIs from trivial
insulators is the existence of topologically protected gapless edge states
that conduct (either charge or spin) even at strong disorder \cite{D1,D2}.
Recently, a novel family of TIs whose helical edge states can be induced
by disorder was discovered \cite
{TAI1,TAI2,TAI6,TAI7,TAI2.5,TAI3,TAI3.5,TAI4,TAI4.1,TAI4.5,TAI4.8,TAI5}.
Explicitly, as disorder strength increases from zero,  the conductance
initially decreases, but then, within a certain region of disorder,
it is quantized at $2e^2/h$ before dropping to zero at stronger disorder
\cite{TAI1,TAI2,TAI6,TAI7,TAI2.5,TAI3,TAI3.5,TAI4,TAI4.1,TAI4.5,TAI4.8,TAI5}.
This (evidently nontrivial) phase of matter is termed as {\it topological
Anderson insulator} (TAI). Specifically, disorder drives topological phase
(TP) transitions by modifying the topological mass and chemical potential
of HgTe quantum well \cite{TAI6}.
Interestingly, TAI in three dimensions has also been identified in a cubic
lattice in which electrons are subject to strong spin-orbit coupling
(SOC) \cite{TAI7}.


So far, TAI was studied mainly in systems
that maintain time-reversal symmetry (TRS) for which the Chern number
vanishes and edge states are helical. In this work we suggest a model for
exposing the (even richer) physics of TAI in systems with broken TRS.
As we show, these systems support multiple chiral edge states, and
display novel TAI phases characterized by nonzero Chern numbers.
Even in the absence of disorder, the model exhibits nontrivial TPs including
the QAH phases with Chern numbers $C=\pm1$ and $\pm2$, and the TRS-broken QSH
phase whose helical edge states are not topologically protected \cite{TRSB}.
Then, adding on-site disorder (of strength $W$), the TAI phases are
identified by re-entrant $e^2/h$ and $2e^2/h$ quantized conductance plateaus.
By using a combination of numerical methods and an effective medium theory
based on the Born approximation \cite{TAI6}, the Chern
number and bulk band gap evolution are evaluated in the $W-E_F$ plane.
Inspection of the nonequilibrium local current distribution (NLCD) in the
nontrivial TAI phases further confirms the conjecture that quantized
conductance plateaus are due to chiral edge states and that transitions between
these plateaus are drivable  by tuning either disorder or Fermi energy.

\section{Model}

In order to study TAI in magnetic systems
that lift spin degeneracy and break TRS, we consider electrons
hopping on a hexagonal lattice subject to SOC, staggered sublattice
potential, antiferromagnetic exchange field \cite{AFM1,AFM2,AFM3,AFM4},
and off-resonant circularly polarized light \cite{L1,L2,L3}.
The tight-binding Hamiltonian reads
\begin{equation}
\begin{split}
H=&t\sum_{\langle ij\rangle}c_{i}^\dagger c_{j} +
\lambda_v\sum_{i} \mu_i c_{i}^\dagger c_{i} + \sum_i\epsilon_i c_i^\dagger c_i \\
&+i\lambda_{so}\sum_{\langle\langle ij\rangle\rangle} \nu_{ij} c_{i}^\dagger s^z c_{j}
+i\lambda_r \sum_{\langle ij \rangle} c_{i}^\dagger (\bm{s}\times\hat{\bm{d}}_{ij})_z c_{j} \\
&+ \lambda_{am}\sum_{i}\mu_i c_{i}^\dagger s^z c_{i}+
i\lambda_l\sum_{\langle\langle ij\rangle\rangle} \nu_{ij} c_i^\dagger c_j.
\end{split}
\label{H1}
\end{equation}
The various symbols in Eq. (\ref{H1}) are defined as follows:
$i$ and $j$ label the lattice sites, while $\langle ij\rangle$
and $\langle\langle ij\rangle\rangle$ stand for nearest
neighbor (NN) and next nearest neighbor (NNN) sites.
$\bm{s}=(s^x, s^y, s^z)$ is the vector of Pauli matrices acting in spin space.
$\nu_{ij}=(2/\sqrt{3})(\hat{\bm{d}}_1\times\hat{\bm{d}}_2)_z=\pm 1$ is defined by
lattice geometry where $\hat{\bm{d}}_1$ and $\hat{\bm{d}}_2$ are two unit
vectors along NN bonds connecting site $i$ to its NNN site $j$,
$\hat{\bm{d}}_{ij}$ is a unit vector connecting two NN sites $i$ and $j$, and
$\mu_i=\pm 1$ denotes A and B sublattices.
The first two terms describe NN hopping and staggered
sublattice potential with respective strengths  $t$ and $\lambda_v$.
The third term represents on-site disorder in which $\{ \epsilon_i \in [
-W/2, W/2] \} $ are i.i.d. The fourth and fifth terms describe the intrinsic
and Rashba SOC with respective strengths $\lambda_{so}$ and $\lambda_r$.
The first five terms respect TRS, and can be viewed as a disordered version
of the Kane-Mele model \cite{T11}. The last two terms break TRS, and are termed
as the antiferromagnetic exchange field \cite{AFM1,AFM2,AFM3,AFM4} and
off-resonant circularly polarized light \cite{L1,L2,L3} with strengths
$\lambda_{am}$ and $\lambda_l$. The last term is named so because it is
originally derived from  the interaction of electrons with a vertically
incident weak circularly polarized light of high frequency ($\gg t/\hbar$). Such that the off-resonant conditions are satisfied and high order effects
can be neglected \cite{L1,L2,L3}. By using the Floquet theory \cite{F1,F2},
the time-dependent problem is transformed to a static problem encoded in the
effective Hamiltonian described by the last term in Eq. (\ref{H1}) \cite
{L1,L2,L3}. Of course, this term may be generated by other means such as by 
staggered magnetic flux in the Haldane model \cite{haldane}.

\section{Clean case: Topological phases}

\begin{figure}
  \includegraphics[width=8.5cm]{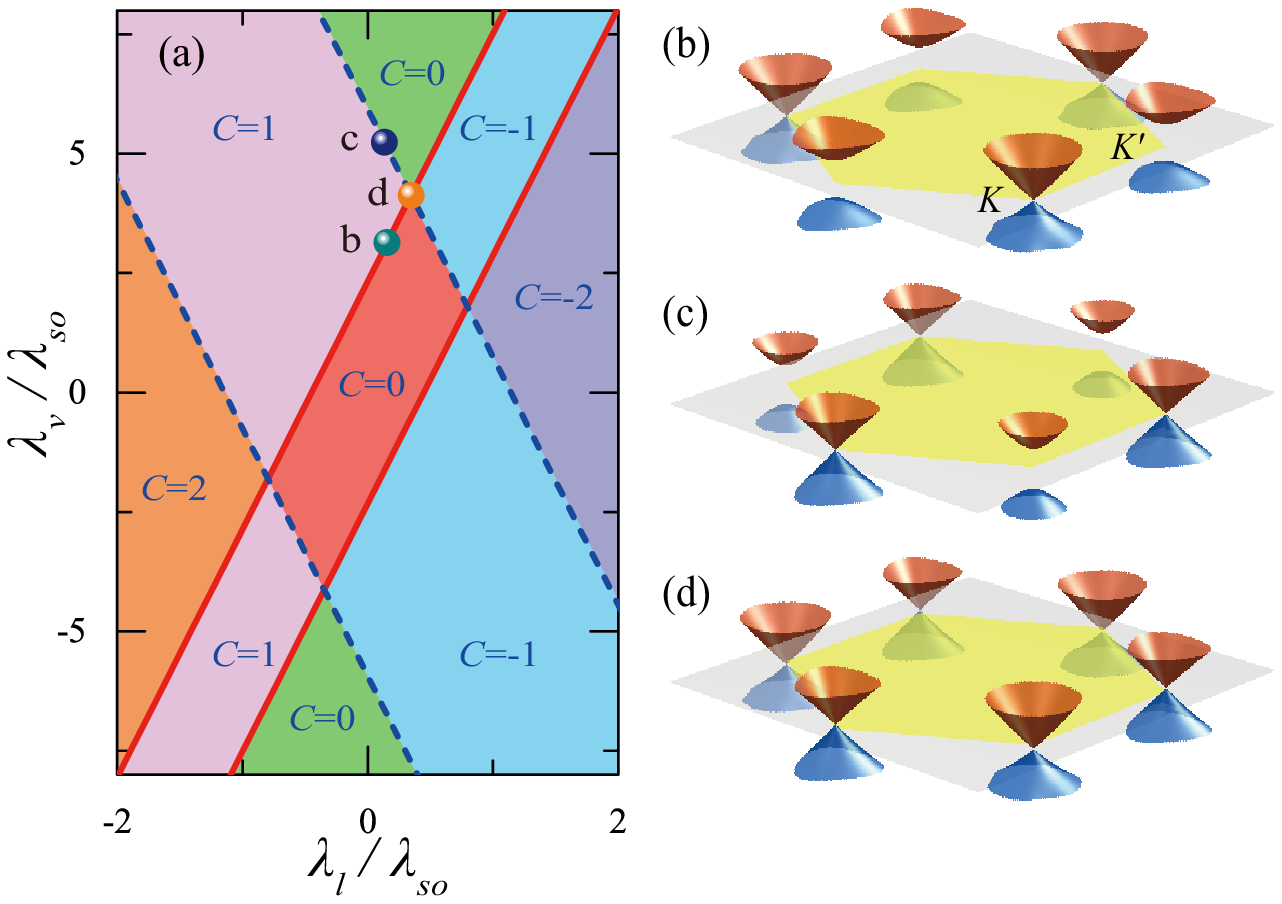}
  \vspace{0.1in}
\caption{(color online) (a) Phase diagram in the $\lambda_l/\lambda_{so}-\lambda
_v/\lambda_{so}$ plane. Red solid and blue dash lines are the TP boundaries
at which the bulk band gap closes at $K$ and $K'$ valleys respectively.
Various TPs with different colors are specified by the Chern number $C$.
The $C=0$ phases are further distinguished by a pair of spin Chern numbers
as $(C_+,C_-)=(0,0)$ marked by green and $(\pm1,\mp1)$ marked by red.
The low-energy band structures near the $K$ and $K'$ valleys (corresponding
to the three TP boundary points labeled as b, c, and d in (a)) are shown
in (b), (c), and (d). }
\label{fig1}
\end{figure}

In the absence of disorder,  this model supports the QAH phases with
$C=\pm1$ and $\pm2$. The Hamiltonian
Eq. (\ref{H1}) is block-diagonized in momentum space as
$H=\sum_{\bm{k}}c_{\bm{k}}^\dagger \mathcal{H}(\bm{k})c_{\bm{k}}$.
In the basis of $\{|\text{A},\uparrow\rangle,|\text{A},\downarrow\rangle,
|\text{B},\uparrow\rangle,|\text{B},\downarrow\rangle\}$,
$\mathcal{H}(\bm{k})$ can be expressed in terms of the Dirac $\Gamma$
matrices \cite{gamma} as
\begin{equation}
\mathcal{H}(\bm{k})=d_0(\bm{k})\text{I}^4+\sum_{a=1}^5 d_a(\bm{k})\Gamma^a +
\sum_{a<b=1}^5d_{ab}(\bm{k})\Gamma^{ab},
\label{H2}
\end{equation}
where $\text{I}^4$ is the $4\times 4$ identity matrix and the
nonvanishing $d_a$ and $d_{ab}$ factors are shown in Ref. \cite{d}.
To be concrete (and without loss of generality), $\lambda_r=\lambda_{am}=0.3t$
and $\lambda_{so}$=$0.2t$ are fixed below while $(\lambda_l,\lambda_v)$
are tunable parameters to realize various TPs (identified by their
corresponding Chern numbers).
TP transitions occur at the closure and reopening of the bulk band gap.
For $\mathcal{H}(\bm{k})$, gap closure and reopening occurs
(at $K$ and $K'$ valley) whenever,
\begin{equation}
2\left|f_\eta\right|-\left|\lambda_\eta\right|=\sqrt{\lambda_\eta^2+9\lambda_r^2},
\label{TB}
\end{equation}
where $f_\eta=\lambda_{am}-3\sqrt{3}\eta\lambda_{so}$ and
$\lambda_\eta=\lambda_v-3\sqrt{3}\eta\lambda_l$, while
$\eta=\pm1$ denotes $K$ and $K'$ valley, respectively.
At half filling, the Chern number is calculated for the valence bands
in the $\lambda_l/\lambda_{so}-\lambda_v/\lambda_{so}$
plane based on the numerical method developed in Ref. ~\cite{C}.
The results are shown in Fig. \ref{fig1}(a) where
TPs are classified by different colors and the Chern numbers.
The TP boundaries of Eq. (\ref{TB}) (red solid and blue
dash lines) separate different TPs as shown in Fig. \ref{fig1}(a).
Crossing points of two different phase boundary lines correspond to
 simultaneous closure of the bulk band gap at two different valleys.
These are verified by the  band structures (Fig. \ref{fig1}(b)-(d))
at three points marked as b, c, d on the TP boundaries as shown
in Fig. \ref{fig1}(a).

\begin{SCfigure*}
  \includegraphics[width=13 cm]{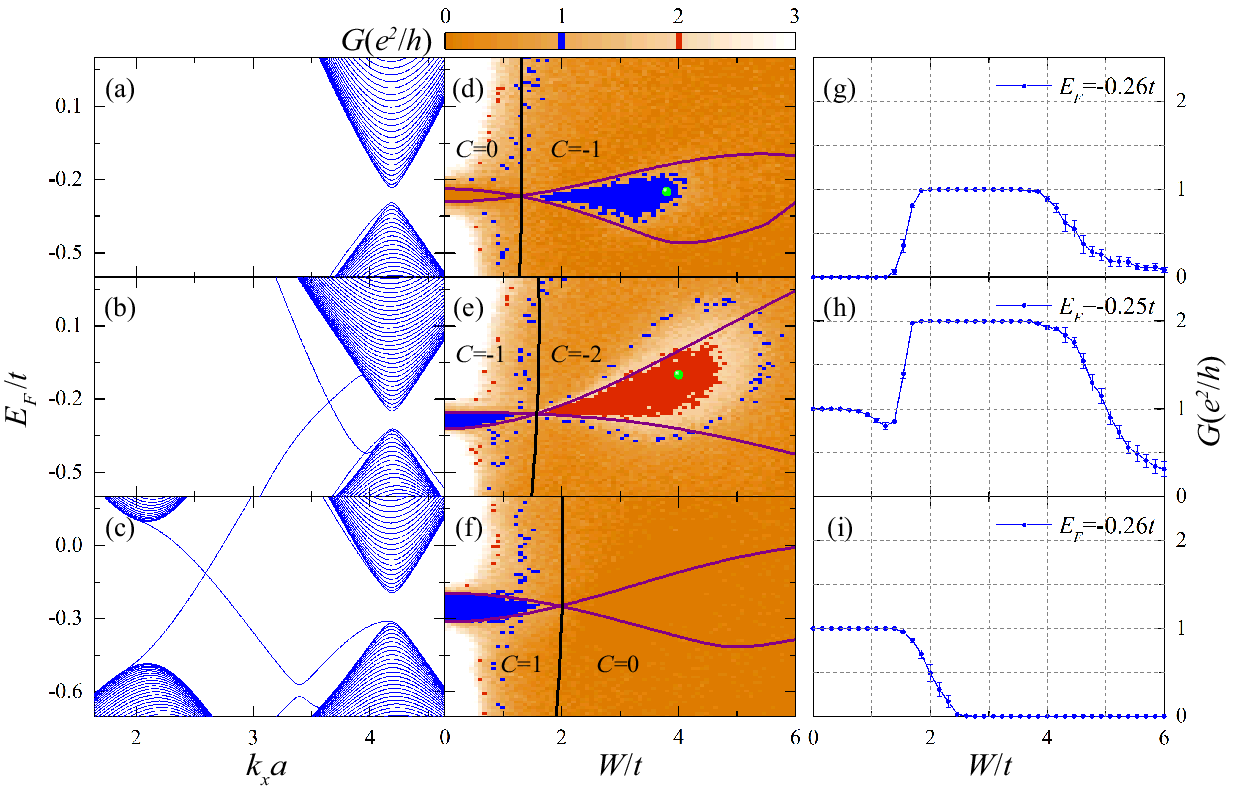}
  \vspace{0in}
\caption{(color online) (a)-(c): Low-energy band structures of the
nanoribbons with zigzag edges in the trivial, QAH1, and QAH2 phases whose
parameters are specified in the text. (d)-(f): Density plot of longitudinal
conductance of the nanoribbons corresponding to (a)-(c). The black
and purple lines are respective TP boundary and bulk band gap evolution
predicted by the Born approximation in the $W-E_F$ plane.
The colorbar is terminated at $G=3$ and $G\geq3$ is represented by white.
(g)-(i): Conductance profiles are displayed as a function
of disorder strength for given Fermi energies within the bulk band gap.}
  \label{fig2}
  \vspace{-0.2in}
\end{SCfigure*}

Two distinct $C=0$ phases exist with different spin Chern numbers $(C_+,C_-)$ 
defined in Ref. \cite{SC1,SC2}: The first, $(C_+,C_-)$=$(0,0)$  (green color 
region in Fig. \ref{fig1}(a)) is a topologically trivial insulator.
The second, $(C_+,C_-)=(\pm1,\mp1)$ (red color region in Fig. \ref{fig1}(a))
is the QSH insulator (strictly protected only if TRS is respected).
When TRS is broken two counter-propagating chiral edge states could annihilate
each other. Nevertheless, when the TRS is only weakly broken, the edge
states that suffer from the backward scattering can still exist, and capable
of transporting charge and spin although not in a perfect (quantized) value \cite{TRSB}.
In other words,  the edge states are not topologically protected,
and, at strong disorder, they will be destroyed \cite{SP}.

In order to elucidate the role of edge states, we consider the model on a
long strip with zigzag edges. The energy spectrum of $\mathcal{H}(\bm{k})$ is  evaluated numerically for $(\lambda_l,\lambda_v)=
(0.25t, 1.8t)$ in the trivial phase, $(0.35t, 1.4t)$ in the
QAH1 phase with $C=-1$, and $(0.03t, 0.7t)$ in  the QAH2 phase with $C=1$.
The corresponding band structures are shown in Fig. \ref{fig2}(a)-(c).
The edge modes are clearly visible for the QAH1 and QAH2 phases.

\section{TAI: Broken TRS and role of disorder}

To study TAI phases within
this model, we elucidate the effect of on-site disorder on the longitudinal
conductance in a two-terminal setup: A long strip of size $200a/\sqrt{3}
\times 1000a$ is connected to two semi-infinite leads at the two ends. Here $a$ is the in-plane
lattice constant. Existence or absence of edge channels due to disorder in different phases can
be clearly seen from the density plot of longitudinal conductance (averaged
over 5 disorder realizations) in the $W-E_F$ plane as shown in Fig.
\ref{fig2}(d)-(f). In the trivial phase, as disorder increases, a $e^2/h$
quantized conductance plateau appears in the bulk band gap as shown in Fig.
\ref{fig2}(d). In the QAH1 phase, as expected, $e^2/h$ quantized conductance
plateau exists within the bulk band gap.
However, it shoots up to $2e^2/h$ above a critical disorder and penetrates into
the conduction band as disorder increases further (see Fig. \ref{fig2}(e)).
In the QAH2 phase, the $e^2/h$ quantized conductance plateau is terminated at
relatively weak disorder as shown in Fig. \ref{fig2}(f).
The corresponding conductance profiles as a function of disorder strength for
given Fermi energies within the bulk band gap are respectively shown in Fig.
\ref{fig2}(g)-(i). The data is obtained by averaging over 100 realizations.
The absence of fluctuation within the quantized conductance plateaus indicates
that they are contributed by topologically protected edge states.

To corroborate this physical interpretation, we analyze the present model
within an effective medium theory based on the Born
approximation in which high order scattering processes are neglected
\cite{TAI6}. In this formalism, the role of disorder is encoded in the
self-energy,
\begin{equation}
\begin{split}
\Sigma(W,E_F) &= \frac{W^2}{12S_{\text{BZ}}}\int_{\text{BZ}} d^2 \textbf
{\textit{k}} \left[ E_F + i0^+ - \mathcal{H}(\textbf{\textit{k}}) \right]^{-1}.
\end{split}
\label{self}
\end{equation}
Here $S_{\text{BZ}}$ is the area of the first Brillouin zone and $\mathcal{H}
(\bm{k})$ is the $4\times4$ Hamiltonian given in Eq. (\ref{H2}).
The self-energy can also be decomposed into the summation of the Dirac
$\Gamma$ matrices. Thus, the effective Hamiltonian in the presence of disorder
is
\begin{equation}
\overline{\mathcal{H}}(\bm{k},W,E_F)=\mathcal{\mathcal{H}}(\bm{k})+\Sigma(W,E_F).
\end{equation}
Due to the violation of time-reversal, spin rotation, inversion, and particle-hole
symmetries, $\Sigma(W,E_F)$ is a full $4\times 4$ matrix (for $W \ne 0$) that
modifies the parameters of $\mathcal{H}(\bm{k})$ and shifts the Dirac
points away from the $K$ and $K'$ valley.

To demonstrate the disorder-induced TP transitions, the Chern number
$C(W,E_F)$ of the effective Hamiltonian $\overline{\mathcal{H}}(\bm{k},W,E_F)$
is evaluated for various $W$ and $E_F$ by using the same method mentioned
early for $\mathcal{H}(\bm{k})$. The results are shown in Fig.
\ref{fig2}(d)-(f) where the TP boundaries are marked by black solid lines
and different TPs are specified by Chern numbers. The disorder-induced
TP transitions are: (i) trivial phase $(C=0)$ $\rightarrow$ QAH phase
$(C=-1)$; (ii) QAH phase $(C=-1)$ $\rightarrow$ QAH phase $(C=-2)$;
(iii) QAH phase $(C=1)$ $\rightarrow$ TRS-broken QSH phase $(C=0)$.
In order to further determine the TAI phases in the $W-E_F$ plane,
the bulk band gap evolution under disorder is evaluated and is represented
by purple lines in Fig. \ref{fig2}(d)-(f). (Recall that the appearance
of edge states requires the Fermi energy  to lie in the bulk band gap).
Indeed,  band inversion occurs at the TP boundary. Thus, for modest disorder,
this effective medium theory yields an adequate description of the various
TAI phases in the $W-E_F$ plane. The peculiarity of these TAI phases is that
they are characterized by nonzero Chern numbers as a result of broken TRS.
In analogy with the distinction between QAH and QSH phases, the TAI phases
found in systems without TRS are distinct from the one studied previously
in systems that respect TRS.

\begin{figure}
  \begin{center}
  \includegraphics[width=8.5 cm]{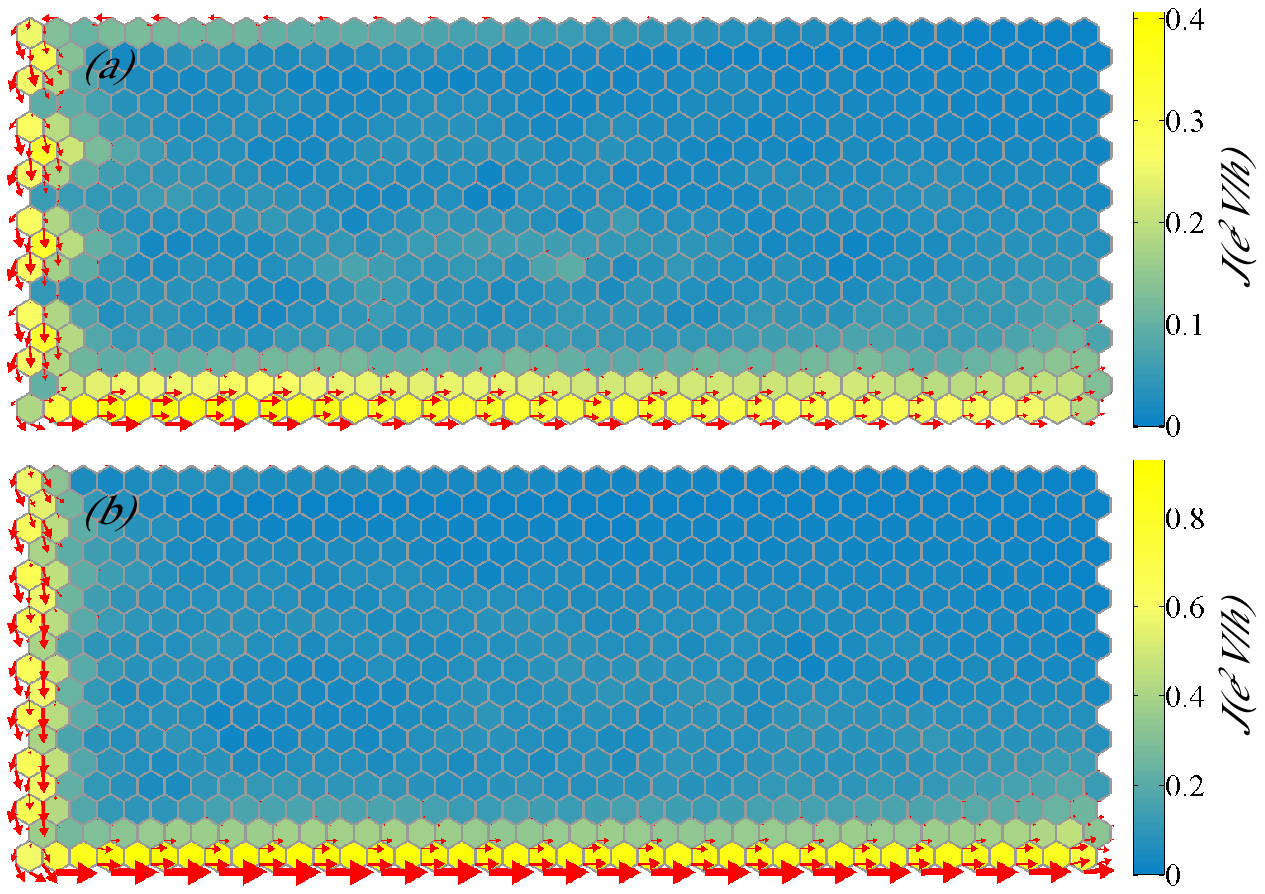}
  \end{center}
  \vspace{-0.2in}
\caption{(color online) The averaged NLCD in nanoribbons for different TAI
phases and with various disorder strengths $W$ and Fermi energies $E_F$
in a two-terminal setup. The color in each hexagon is the average value
of the NLCs at six corners. The red arrows denote current direction at
each lattice site. The arrow size is proportional to the magnitude of NLC.
$eV$ denotes the chemical potential difference between the source and drain.
(a) $W=3.8t$ and $E_F=-0.25t$ (marked by green spot in Fig. \ref{fig2}(d)).
(b) $W=4t$ and $E_F=-0.1t$ (marked by green spot in Fig. \ref{fig2}(e)).}
  \label{fig3}
  \vspace{-0.2in}
\end{figure}

\section{Nonequilibrium local current distribution (NLCD)}

To further
substantiate the assertion that the re-entrant quantized conductance plateaus in Fig.
2(d) and (e) originate from the robust chiral edge states, we study the NLCD
(averaged over 1000 ensembles). In the two-terminal setup specified above,
the strip size is set as $26a/\sqrt{3} \times 40a$ here. Consider first the disorder
induced $e^2/h$ quantized conductance plateau shown in Fig. \ref{fig2}(d).
For $(W,E_F)=(3.8t,-0.25t)$ (green dot in Fig. \ref{fig2}(d)) that is in the
$C=-1$ TAI phase, the NLCD is shown in Fig. \ref{fig3}(a). Apparently, the
nonequilibrium local currents (NLCs) are strongly localized at the bottom edge.
The NLCs at the top edge are suppressed because the
charge motion is directed against the source-drain bias voltage.
Similar NLCD is found in the $2e^2/h$ quantized conductance plateau.
For $(W,E_F)=(4t,-0.1t)$ (marked by green spot in Fig. \ref{fig2}(e))
within the $C=-2$ TAI phase, the NLCD is shown in Fig. \ref{fig3}(b).
As expected, the edge current is roughly twice higher than that
encountered in Fig. \ref{fig3}(a) since the Chern number is doubled.
Namely, the number of edge channels is doubled.
The fact that the two edge modes propagate in the same direction at
the same edge indicates that the edge states are chiral, not helical.
Thus, our results pertaining the NLCD are consistent with the
conclusions obtained within the Born approximation.

\section{Discussion}

So far, the TAI was observed and analyzed in systems that respect TRS. In these 
systems, the Chern number is strictly zero. Thus only trivial and QSH phases can 
exist, and there are either zero or two Kramer degenerate helical edge channels. 
Consequently, the conductance can only jump from $0$ to $2e^2/h$ or vice versa. 
It was not clear from previous studies how the topologically protected edge 
channels are created and destroyed by disorder in systems without TRS, where 
the TAI phases can have nonzero Chern numbers and support chiral edge states. 
In fact, the appearance of robust multiple chiral edge modes in our model is 
striking and nontrivial within our current understanding of TAI \cite{D2}.
The present work shows the edge channels can be created and/or destroyed
one-by-one, resulting in multiple quantized conductance plateaus.
Furthermore, previous works can only be realized in non-magnetic materials.
Our model could be realized in magnetic materials.
As for the experimental detection, the TAI in systems without TRS can support
both even and odd numbers quantized conductance plateaus in units of $e^2/h$.
However, only even number quantized conductance is permitted in TAI with
TRS studied previously. For QSH states in the present case where the TRS is
broken, helical edge states are not topologically protected as expected.
However, these edge states, suffering the backward scattering, can still
exist at weak disorder and transport charge and spin \cite{TRSB,SP}.

\section{Summary}

We present a tight-binding model on a hexagonal lattice that
breaks both time-reversal and spin rotation symmetries. It displays several TPs,
including the QAH phases with $C=\pm1$ and $C=\pm2$, as well as the TRS-broken QSH phase.
In the presence of on-site disorder, the longitudinal conductance within
various TPs is displayed in the $W-E_F$ plane and exposes new TAI phases.
They are characterized by nonzero Chern numbers, and can be experimentally
identified from the re-entrant $e^2/h$ and $2e^2/h$ quantized conductance plateaus.
Disorder-induced TP transitions are marked by jumps of the
quantized conductance from 0 to $e^2/h$ and from $e^2/h$ to $2e^2/h$.
An effective medium theory based on the Born approximation adequately encodes
all transitions among these TAI phases.
Analysis of the NLCD further confirms the interpretation that
quantized conductance plateaus originate from  chiral edge states,
similar to the QAH and quantum Hall effects.

\section{Acknowledgment}

This work is supported by NSFC of China grant (11374249)
and Hong Kong RGC grants (163011151 and 605413). The research of Y.A is
partially supported by Israeli Science Foundation grant 400/2012.


\begin{thebibliography}{99}
\bibitem{gamma} The Dirac $\Gamma$ matrices are defined as: $\Gamma^{(1,2,3,4,5)}=(\tau^x \text{I}^2,\tau^z \text{I}^2,\tau^y s^x,\tau^y s^y, \tau^y s^z)$ and $\Gamma^{ab}=[\Gamma^a,\Gamma^b]/(2i)$ \cite{T11}, where $
\tau^k$ and $s^k$ are Pauli matrices acting in sublattice and spin spaces.
\bibitem{d} The nonvanishing $d_a$ and $d_{ab}$ factors in Eq. (\ref{H2}) are: $d_1=t(1+2\cos x\cos y)$, $d_2=\lambda_l(2\sin2x - 4\sin x\cos y)+\lambda_v$, $d_3=\lambda_r(1-\cos x\cos y)$, $d_4=-\sqrt{3}\lambda_r\sin x\sin y$, $d_{12}=-2t\cos x\sin y$, $d_{15}=\lambda_{so}(2\sin2x - 4\sin x\cos y)+\lambda_{am}$, $d_{23}=-\lambda_r\cos x\sin y$, $d_{24}=\sqrt{3}\lambda_r\sin x\cos y$, where $x=k_xa/2$ and $y=\sqrt{3}k_ya/2$.
\end{thebibliography}
\end{document}